\documentclass[prl,aps,twocolumn,showpacs,floatfix]{revtex4-1}%
\usepackage{amsmath}
\usepackage{amsfonts}
\usepackage{amssymb}
\usepackage{graphicx}
\usepackage{bm}

\begin{document}

\title{London penetration depth and strong pair-breaking in iron-based superconductors}

\author{R.~T.~Gordon}
\author{H.~Kim}
\author{M.~A.~Tanatar}
\author{R.~Prozorov}
\author{V.~G.~Kogan}
\affiliation{Ames Laboratory and Department of Physics \& Astronomy, Iowa State University, Ames, IA 50011}



\begin{abstract}
The low temperature variation of the London penetration depth for a number of iron-pnictide and iron-chalcogenide superconductors is nearly quadratic,
$\Delta \lambda(T) = \beta T^n$ with $n\approx 2$. The coefficient in this dependence shows a robust scaling, $\beta \propto 1/T_c^3$
across  different families of these materials. We associate the scaling with a strong pair-breaking. The same mechanism
have recently been suggested  to explain the  scalings of the specific heat jump, $\Delta C \propto T_c^3$ \cite{BNC}, and of the
slopes of the upper critical field, $dH_{c2}/dT\propto T_c$ in these materials \cite{K2009}. This suggests  that thermodynamic
and electromagnetic properties of the iron-based superconductors can be described within a strong pair-breaking scenario.

\end{abstract}

\maketitle

Due to the unique electronic structure and, most likely, unconventional pairing mechanism, iron-based superconductors
exhibit a number of uncommon properties. It has recently been reported \cite{BNC} that across the whole family of  iron-pnictides
the specific heat jump, $\Delta C$, at the critical temperature $T_c$  shows an extraordinary scaling   $\Delta C \propto T_c^{3}$,
whereas in conventional s-wave materials $\Delta C\propto  T_c $. According to Ref.\,\onlinecite{K2009}, this unusual scaling is
caused by a strong pair-breaking  in materials with anisotropic order parameters; both transport and magnetic scattering in such
materials suppress $T_c$ and there are  plenty of reasons for magnetic pair breaking in iron-based superconductors. Another
consequence of this model, proportionality of slopes of the upper critical field $[dH_{c2}/dT]_{T_c}$ to $T_c$, has also been shown
to hold for the data available \cite{K2009}. In this work we show that the same idea can be applied to the low temperature behavior of the London penetration, $\Delta\lambda=\lambda(T)-\lambda(0)$,
where the pair-breaking results in

\begin{equation}
\Delta\lambda \propto  T^2/T^3_c \,,
    \label{dlambda}
\end{equation}

Despite some initial disagreements in experimental reports, most precision measurements of the in-plane London penetration
depth of  iron-based superconductors had found the power-law behavior, $\Delta\lambda (T)\propto T^n$ with $n \approx 2$
\cite{BaCo122PRL,BaCo122PRB,PDPhysicaC,1111PRL,BaK122PRB,Hashimoto2009a,Luan2009};
for some compounds $n \approx 1$ is claimed \cite{Fletcher2009,Hicks2009,Hashimoto2009}. Commonly, a non-exponential behavior is
taken as evidence of unconventional  order parameter, possibly having a nodal gap structure  \cite{Vorontsov2009,Hashimoto2009a,BaCo122PRB,BaK122PRB}.
However, such a direct correspondence between the nodes and the  exponent $n$ should exist  only in clean materials. As a rule, scattering breaks this
elegant connection. E.g., for   d-wave superconductors, the linear low $T$ dependence of $\lambda$ in the clean case changes to $T^2$ in the
presence of moderate scattering \cite{HirGold}. In fact, connection between the power-law behavior of $\Delta \lambda(T)$ and scattering in pnictides had been
suggested \cite{Vorontsov2009,Hashimoto2009a,BaCo122PRB,BaK122PRB}. The symmetry of the order parameter $\Delta$ in multi-band iron-pnictides is not yet determined with certainty, however, many favor  the  $s_{\pm}$ structure \cite{Mazin-Schm,Junhwa}. The Fermi surface (FS) average of the order parameter in this model  $\langle \Delta\rangle \ll \Delta_{max}$. We then expect the penetration depth to  behave like a  ``dirty" d-wave, i.e., to show the low-temperature variation $\propto T^2$.

The samples were plate-like single crystals with typical dimensions 1 x 1 x 0.2 $mm^3$. Details of sample synthesis and characterization can be found
elsewhere \cite{Ni2008,PCC2009,Fang2008}. The penetration depth measurements were performed with a self-resonating tunnel diode oscillator. Diamagnetic response of the sample causes shift of the resonant frequency, $\Delta f =  - G \chi(T)$, where $\chi(T)$ is magnetic
susceptibility determined by $\lambda(T)$ in the Meissner state, $-4\pi\chi(T)=[1-(\lambda/R)\tanh(R/ \lambda)]$. The calibration constant $G = f_0V_s/2V_c(1-N) $ is measured directly by extracting the sample from the coil at the lowest temperature. Here $f_0 \approx 14$ MHz is the empty resonator frequency, $V_s$ and $V_c$ are the sample and coil volumes, and $N$ is the demagnetization factor. Details of measurements and of data
analysis are described elsewhere \cite{Prozorov00,Prozorov06}

\begin{figure}[tb]
\includegraphics[width=8.5cm]{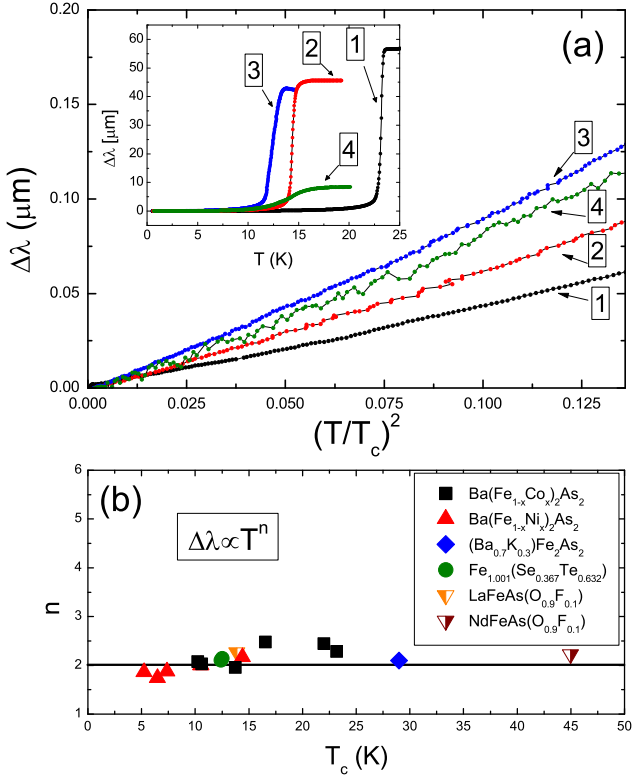}
\caption{(Color online) (a) $\Delta\lambda$  versus $(T/T_c)^2$ for
Ba(Fe$_{0.942}$Co$_{0.058}$)$_2$As$_2$ marked by (1),
Ba(Fe$_{0.941}$Ni$_{0.059}$)$_2$As$_2$ (2),
Fe$_{1.001}$Se$_{0.367}$Te$_{0.632}$  (3), and
LaFeAsO$_{0.9}$F$_{0.1}$  (4). Inset: (a) $\Delta\lambda$ in
the full temperature range. (b) Fitted exponent $n$ in $\Delta\lambda \propto T^n$.}
\label{fig1}
\end{figure}

Figure  \ref{fig1}(a) shows the linear behavior of
$\Delta\lambda$ versus ($T/T_c)^2$  for      $T<T_c$/3  in
iron-based compounds with $T_c$ varying from $\approx$ 12 to 23\,K;
the data are  from Refs.\,[\onlinecite{BaCo122PRL, BaCo122PRB,
PDPhysicaC, 1111PRL, BaK122PRB, FeSePhysicaC}].  The  exponent $n$
in $\Delta\lambda \propto T^n$ extracted by fitting the low
temperature data is shown for six compounds   in
Fig.\,\ref{fig1}(b). We see  that  $\Delta\lambda(T)\propto T^2$
holds for the (AE)(Fe$_{1-x}$TM$_x$)$_2$As$_2$ (``122"),
(RE)FeAs(O$_{1-x}$F$_x$) (``1111"), and FeTe$_{1-x}$Se$_x$ (``11") families; here AE stands for an alkali  earth element, TM
for a transition metal, RE for a rare earth. Thus,  the  four lines shown in Fig.\,\ref{fig1}(a)
are not  merely for different doping levels of the same compound,  but
rather they belong to four different families of   the  iron-based
materials. This universal behavior has prompted us to look for a
universal cause; we offer below a strong pair-breaking as such a
cause.

The theoretical tool we employ, the quasiclassical version of
the weak-coupling Gor'kov theory, holds for a general
anisotropic Fermi (F) surface  and for any gap symmetry \cite{E}.
The formalism in the form convenient for our purpose is outlined in
Ref.\,[\onlinecite{K2009}\;  we refer readers  to this work for
details. The theory is formulated in terms of   functions $f({\bm
r},{\bm k}_F,\omega),\,\,  f^{+} $, and $g$ which originate from
Gor'kov's Green's functions and are normalized by $g^2+ff^+=1$;
the  Matsubara frequencies are $\omega=\pi T(2\nu+1)$ with an
integer $\nu$ and $\hbar=k_B=1$. The order parameter  is taken in
the form
$ \Delta ({\bm r},{\bm k}_F)=\Psi ({\bm  r},T)\,
\Omega({\bm k_F})$ where  $ \Omega({\bm  k}_F)$   describes the
variation of $\Delta$ along the F-surface and is conveniently
normalized so that the average  over the whole F-surface  $ \langle
\Omega^2  \rangle=1$. Hence, the model is a BCS-type weak-coupling
approach providing   a qualitative description at best.

The  scattering in the Born approximation is characterized
by two scattering times, the transport   $\tau $   responsible for
the normal conductivity and  $\tau_m$ for processes breaking the
time reversal symmetry (e.g.,   spin-flip):
\begin{equation}
 1/\tau_\pm = 1/\tau \pm 1/\tau_m  \,.
  \label{taus}
  \end{equation}
Commonly,  two dimensionless parameters are used:
  \begin{equation}
\rho= 1/2\pi T_c\tau \qquad {\rm and}\qquad
\rho_m= 1/2\pi  T_c\tau_m\,,
  \label{rhos}
\end{equation}
or equivalently $\rho_\pm= \rho\pm \rho_m$. This is of course
 a gross simplification. For multi-band F-surfaces one may need
more parameters for various intra- and inter-band processes, which are hardly controllable and their number
is too large for  a useful theory.  Our model  is amenable for analytic work and may prove helpful, the simplifying assumptions notwithstanding.

It is well-known that the formal  scheme of the seminal Abrikosov-Gor'kov (AG) work on magnetic impurities \cite{AG} applies to various situations with different pair-breaking causes, not necessarily the AG spin-flip scattering \cite{Maki}. In each particular situation, the  parameter
  $\rho_m$ must be properly defined. Here, without specifying the
pair breaking  mechanism, we apply  the AG approach to show that
the pair-breaking accounts for our data on the low temperature
$\lambda(T)$   along with the earlier reported behavior of
$H_{c2}$   slopes at $T_c$ and of the quite unusual dependence of the specific heat jump on $T_c$.

Evaluation of $\lambda(T;\tau,\tau_m)$ for arbitrary $\tau$'s
and arbitrary anisotropy of $\Delta$ is difficult analytically.
However, for a strong $T_c$ suppression, the problem is manageable.
Within the microscopic theory,   penetration of weak magnetic
fields into superconductors is evaluated by first solving for the
unperturbed zero-field  state and then treating   small fields as
perturbations. It was shown by AG \cite{AG} that  for strong
pair-breaking the formalism for the derivation of the
Ginzburg-Landau equations near $T_c$ applies at all temperatures.
Within the Eilenberger approach this means that $f\ll 1$ and
$g\approx 1-ff^+/2$ at all temperatures.  The calculation then
proceeds in a  manner similar to that near $T_c$.

Within a two-band model for iron-based materials, the order
parameter is believed to have a $\pm s$ structure
\cite{Mazin-Schm}, so that $\langle\Delta\rangle \ll
|\Delta_{max}|$ \cite{Junhwa}.  The problem is considerably
simplifies if one assumes $\langle\Delta\rangle=0$; we use this
assumption and expect  the model  to hold at least qualitatively.
In the zero-field state, we look for solutions of Eilenberger
equations as $f_0=f^{(1)}+f^{(2)}+...$ where $ f^{(1)}\sim\Delta $,
$f^{(2)}\sim\Delta^2$, etc.  The Eilenberger equation for $f$ then
yields \cite{K2009}:
\begin{eqnarray}
f_0&=&\frac{\Delta}{\omega_+}
 +\frac{\Delta}{2\omega_+^3}\left(\frac{
\langle\Delta^2\rangle}{2\tau_+\omega_+}-\Delta^2\right) +
{\cal O}(\Delta^5),
\label{f4}
\end{eqnarray}
where $\omega_+ = \omega+1/2\tau_+$. One can see that even at
low temperatures $f_{0,max} \sim \tau_+T_c \sim 1/\rho_+ \ll 1$
because for strong pair-breaking   $T_c\to 0$. This is a
quasiclassical justification for the  AG statement that $f\ll 1$ at
all $T$'s.

The $T$ dependence of $\Delta$ (or $\Psi $) is obtained with
the help
of the self-consistency equation (or the ``gap equation"). For a strong pair-breaking, this
equation takes the form \cite{K2009}:
\begin{equation}
\frac{\Psi(1-t^2)}{12\pi T\rho^2_+}  = \sum_{\omega
>0}^{\infty}\left(\frac{\Psi}{\omega^+}  -\Big\langle \Omega \, f
\Big\rangle\right)  \,.
\label{self-cons1}
\end{equation}
Substituting here $f$ of Eq.\,(\ref{f4}), we obtain  the order parameter in the field-free state:
\begin{eqnarray}
 \Psi^2 =\frac{2\pi^2(T_c^2 -T^2)}{3\langle\Omega^4\rangle-2  } \,;
\label{Psi}
\end{eqnarray}
  this reduces to the AG form for $\Omega=1$.

  We can now consider the  response to a small current
\begin{eqnarray}
{\bm  j}=-4\pi |e|N(0)T\,\, {\rm Im}\sum_{\omega >0}\Big\langle {\bm v}g\Big\rangle\,;
\label{eil5}
\end{eqnarray}
$N(0)$ is the density  of states at the F-level per one spin. Weak
supercurrents  leave the order parameter modulus unchanged, but
cause the condensate to acquire an overall phase $\theta({\bm  r})$.
We then look for   perturbed solutions as:
\begin{eqnarray}
\Delta  = \Delta _0 \, e^{i\theta},\,\,\,\,\,
f =(f_0  +f_1)\,e^{i\theta},\nonumber\\
f^{+} =(f_0  +f_1^+ )e^{-i\theta},\,\,\,\,\,
g =g_0 +g_1             ,
\label{perturbation}
\end{eqnarray}
where the subscript 1 denotes small corrections to the uniform
state $f_0,g_0$.  In the London limit, the only coordinate
dependence is that of the phase $\theta$, i.e., $f_1 ,g_1 $
are ${\bm  r}$ independent. The Eilenberger equations  provide the
corrections among which we need only $g_1$:
\begin{equation}
g_1=\frac{i  f_0^2 {\bm v}{\bm P}}{2 \omega_+ }= \frac{i
 \Delta^2 {\bm v}{\bm P}}{2 \omega_+^3 }\,.
\label{g1a}
\end{equation}
 see \cite{K2009}. Here ${\bm  P}= \nabla\theta+ 2\pi{\bm
A}/\phi_0\equiv 2\pi\, {\bm  a}/\phi_0$ with the ``gauge
invariant
vector potential" ${\bm  a}$.

We now substitute $g_0+g_1$ in Eq.\,(\ref{eil5}) and compare
the result with   $4\pi
j_i/c=-(\lambda^2)_{ik}^{-1}a_k$ to obtain:
\begin{equation}
 (\lambda^2)_{ik}^{-1}= \frac{8\pi^2 e^2N(0)T_c }{ c^2 }
\Big\langle v_iv_k \Omega
 ^2\Big\rangle
\Psi^2\sum_{\omega>0} \frac{1}{\omega^3_+}\,.
    \label{lambda-2}
\end{equation}
The sum here is expressed in terms of the polygamma function:
\begin{equation}
\sum_{\omega>0} \frac{1}{\omega^3_+}=-\frac{1}{16\pi^3T^3}
 \psi^{\prime\prime}\left(\frac{\rho^+ }{2t}+\frac{1}{2}\right) \approx \frac{\tau_+^2}{\pi T}\,,
    \label{summ}
\end{equation}
 where $\rho_+\gg 1$ has been used. Taking into account
Eq.\,(\ref{Psi}), one obtains:
\begin{equation}
 (\lambda^2)_{ik}^{-1}= \frac{16\pi^3 e^2N(0)k_B^2\tau_+^2 }{ c^2 \hbar^2(3\langle\Omega^4\rangle-2)}
\Big\langle v_iv_k \Omega
 ^2\Big\rangle(T_c^2-T^2)\,
    \label{lambda_gapless}
\end{equation}
\noindent in common units. It is now easy to obtain
 the low $T$ behavior of
$\Delta\lambda_{ab}=\lambda_{ab}(T)-\lambda_{ab}(0) $ for a
uniaxial material:
\begin{eqnarray}
\Delta\lambda_{ab}=\eta\,\frac{T^2}{T^3_c},\, \,\,\,
\eta = \frac{c\hbar}{8\pi k_B \tau_+}\sqrt{\frac{
 3\langle\Omega^4\rangle-2 }{ \pi
e^2N(0)\langle v_a^2 \Omega
 ^2 \rangle } }.
    \label{dlambda}
\end{eqnarray}
We stress  that $\tau_+$ here is close to the critical value
for which $T_c\to 0$. One readily obtains for $T=0$,
\begin{eqnarray}
 \lambda_{ab}(0)= 2\eta/T _c \,.
     \label{lambda(0)}
\end{eqnarray}
Note: Eqs.\,(\ref{dlambda}) and (\ref{lambda(0)}) are derived
for $\langle\Omega\rangle\approx 0$. One can show that they  hold
for
$\langle\Omega\rangle\ne 0$ as well with, however, different
coefficient $\eta$. We do not provide  here this cumbersome
calculation.

\begin{figure}[tb]
\includegraphics[width=8.5cm]{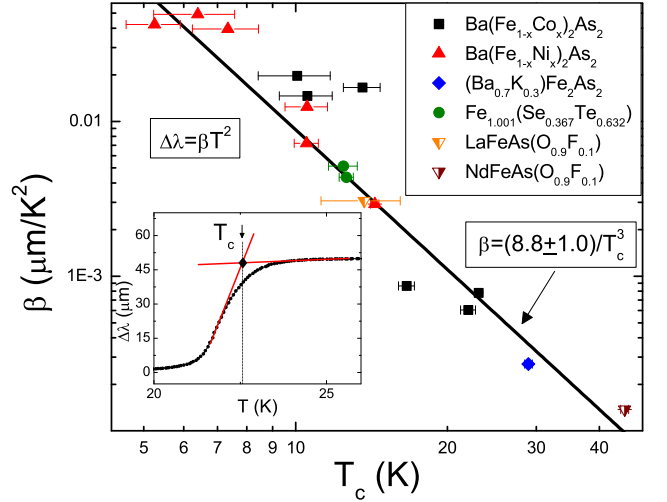}
\caption{(Color online) The factor $\beta$ obtained in fitting of data to $\Delta\lambda=\beta T^2$  plotted versus $T_c$ on a log-log scale. The solid line is a fit to $\beta=\eta/T_c^3$, motivated by Eq.\,(\ref{dlambda}) for a strong pair-breaking.}
\label{fig2}
\end{figure}

To examine the predicted scaling behavior, the  factor $\beta$
 in  $\Delta\lambda=\beta T^2$  was obtained  by fitting the low
temperature   $\Delta\lambda$  for the same   122, 1111 and 11
compounds  of Fig.\,\ref{fig1}  with $\beta$ being the only fitting
parameter. The   $\beta$'s are plotted in the main panel of
Fig.\,\ref{fig2}  versus $T_c$.  The  error bars on this graph
reflect the fact that each sample studied has a certain transition
width. The inset of Fig.\,\ref{fig2} shows the convention adopted
here for $T_c$ determination. The uncertainty of $T_c$  is the
dominant source of error in   determination of  $\beta$.  According
to the strong pair breaking scenario, $\beta=\eta/T_c^3$.  To
compare  experiment with theory,  $\beta$   is plotted on a log-log
scale in the main frame of Fig.\,\ref{fig2} along with the line
$\beta=(8.8 \pm 1.0)/T_c^3$   obtained by  fitting the data.
Moreover, by substituting $v\sim 10^7\,$cm/s and $N(0)\sim 10^{33}\,$
erg$^{-1}$cm$^{-3}$ in Eq.\,(\ref{dlambda}) we roughly estimate
$\tau^+\sim  3\times 10^{-14}\, $s; this value corresponds to
parameter $\rho^+\approx 5$ for $T_c= 40 \,$K and to larger values
for lower $T_c$'s, an observation consistent with the major model
assumption of $\rho^+\gg 1$.
 The degree to which the experimental values follow the theory
 is remarkable, a substantial scatter of the data points
notwithstanding.

The scalings of Eqs.\,(\ref{dlambda}) and  (\ref{lambda(0)})
are obtained for a strong pair-breaking materials with the order
parameter obeying $\langle\Omega \rangle\approx 0$. These
conditions are likely to be satisfied in underdoped
high-$T_c$ cuprates since underdoped materials are clearly disordered
and the d-wave order parameter is suppressed by any scattering.
Indeed, the surface resistance \cite{Bonn} and  optical data
\cite{Homes}  for    YBa$_2$Cu$_3$O$_{6+x}$ samples
with $T_c$ varying from 3 to 17\,K  show
$1/\lambda_{ab}^2(0)\propto T_c^2$ in agreement with
Eq.\,(\ref{lambda(0)}).  This behavior differs from ``Uemura
scaling" $1/\lambda_{ab}^2(0)\propto T_c$ \cite{Uemura89}.

To our knowledge there is not yet sufficient data on
$\lambda(0)$ for the iron-based materials to verify the scaling
(\ref{lambda(0)}). Similarly, we are not aware of a  data set to
check the scaling $H_{c1} \propto T_c^2$ which follows from
Eq.\,(\ref{lambda(0)}).

We would like to stress that the penetration depth scalings
discussed here as well as those for the the specific heat jump and
for the slopes of $H_{c2}(T)$ described in Ref.\,[\onlinecite{K2009}]
are approximate by design since their  derivation  involves a
number of simplifying assumptions. Still they are robust in showing
that the pair-breaking is an important factor in superconductivity
of iron-pnictides.

Many questions still remain; for example,  why the Co doped 122 compounds deviate substantially from the general scaling behavior
shown in Fig.\,\ref{fig2}, see also  Ref.\,[\onlinecite{K2009}]. Another problem to address is how to reconcile the strong
pair-breaking, which in the {\it isotropic} case leads to gapless superconductivity \cite{AG}, with the in-plane  thermal conductivity
data showing $\kappa (0)=0$ \cite{Luo2009,Tanatar2009}. At this point, we can say that (a) the strong pair-breaking model for {\it
anisotropic} order parameters  states  that the {\it total} density of states $N(\epsilon)$ integrated over all pockets  of
the  Fermi surface is finite   at zero energy \cite{K2009};  this
does not exclude a possibility that $N=0$ for some parts on the
Fermi surface. And (b): in this work we are interested in the
superfluid density $\propto 1/\lambda^2$ which depends only on the
Fermi surface average $\langle\Delta\rangle$ so that our results
are less sensitive to the $\Delta$ behavior on a particular set of
directions (e.g., those in the $ab$ plane). The same qualitative
argument  shows that our scalings do not contradict the in-plane
ARPES data
\cite{ARPES}.

To conclude, analysis of the low-temperature behavior of the
London penetration depth shows that a strong pair-breaking is
likely to be responsible for the nearly universal  temperature
dependence $\Delta\lambda_{ab}\propto  T^2/T^3_c$,  along with
earlier reported $\Delta C\propto T_c^{-3}$ and
$[dH_{c2}/dT]_{T_c}\propto T_c$, in nearly all  iron-based
superconductors.

We thank S. L. Bud'ko, P. C. Canfield, A. Chubukov, K. Hashimoto, C. Martin,
Y. Matsuda, K. A. Moler, H.-H. Wen, Zh. Mao for helpful discussions. Work at the Ames Laboratory was
supported by the Department of Energy - Basic Energy Sciences under Contract No. DE- AC02-07CH11358. R. P. acknowledges support of
Alfred P. Sloan Foundation.

 \end{document}